\documentclass[conference]{IEEEtran}
\IEEEoverridecommandlockouts
\usepackage{cite}
\usepackage{amsmath,amssymb,amsfonts}
\usepackage{algorithmic}
\usepackage{graphicx}
\usepackage{textcomp}
\usepackage{xcolor}
\def\BibTeX{{\rm B\kern-.05em{\sc i\kern-.025em b}\kern-.08em
    T\kern-.1667em\lower.7ex\hbox{E}\kern-.125emX}}

\usepackage{hyperref}
\usepackage{breakurl}
\hypersetup{
    breaklinks=true,
    colorlinks=true,
    urlcolor=blue
}

\begin{document}

%
\makeatletter
\newcommand{\linebreakand}{%
   \end{@IEEEauthorhalign}
   \hfill\mbox{}\par
  \mbox{}\hfill\begin{@IEEEauthorhalign}
}
\makeatother
    
\title{Innovative Approaches to Teaching Quantum Computer Programming and Quantum Software Engineering\\

\thanks{This work has been supported by the Academy of Finland (project DEQSE 349945) and Business Finland (project TORQS 8582/31/2022). Also, this work has been partially funded by the EU "Next GenerationEU/PRTR", by the Ministry of Science, Innovation and Universities (projects PID2021-1240454OB-C31 - AEI/10.13039/501100011033/FEDER, TED2021-130913B-I00, PDC2022-133465-I00, and RED2022-134148-T); by the Regional Ministry of Economy, Science and Digital Agenda of the Regional Government of Extremadura (GR21133); and by EU under the Agreement 101083667 of the Project "TECH4E -Tech4effiency EDlH"}
\thanks{Majid Haghparast and Tommi Mikkonen are with the Faculty of Information Technology, University of Jyväskylä, Jyväskylä, Finland. (corresponding author e-mail: majid.m.haghparast@jyu.fi)}
\thanks{Jose Garcia-Alonso, Enrique Moguel, and Juan M. Murillo are with the Universidad de Extremadura, Cáceres, Spain.}

}

\author{\IEEEauthorblockN{Majid Haghparast}
\IEEEauthorblockA{
\textit{University of Jyväskylä}\\
Jyväskylä, Finland \\
majid.m.haghparast@jyu.fi}
\and
\IEEEauthorblockN{Enrique Moguel}
\IEEEauthorblockA{
\textit{University of Extremadura}\\
Cáceres, Spain \\
enrique@unex.es}\\
\and
\IEEEauthorblockN{Jose Garcia-Alonso}
\IEEEauthorblockA{
\textit{University of Extremadura}\\
Cáceres, Spain \\
jgaralo@unex.es}\\
\linebreakand
\IEEEauthorblockN{Tommi Mikkonen}
\IEEEauthorblockA{
\textit{University of Jyväskylä}\\
Jyväskylä, Finland \\
tommi.j.mikkonen@jyu.fi}
\and
\IEEEauthorblockN{Juan Manuel Murillo}
\IEEEauthorblockA{
\textit{University of Extremadura}\\
Cáceres, Spain \\
juanmamu@unex.es}

}

\maketitle

\begin{abstract}
Quantum computing is an emerging field that promises to revolutionize various domains, such as simulation optimization, data processing, and more, by leveraging the principles of quantum mechanics. This paper outlines innovative pedagogical strategies developed by university lecturers in Finland and Spain for teaching quantum computer programming and quantum software engineering. Our curriculum integrates essential tools and methodologies such as containerization with Docker, Qiskit, PennyLane, and Ocean SDK to provide a comprehensive learning experience. The approach consists of several steps, from introducing the fundamentals of quantum mechanics to hands-on labs focusing on practical use cases. We believe quantum computer programming is an important topic and one that is hard to teach, so having a teaching agenda and guidelines for teaching can be of great help.
\end{abstract}


\section{Introduction}
\IEEEPARstart{Q}{uantum} computer programming course is an advance level course usually discussed in M.Sc. or Ph.D. level. There is an increased need to develop novel methods to teach quantum computer programming and quantum software engineering. As university professors and lecturers in Finland and Spain, we have developed innovative approaches to teaching quantum computer programming and quantum software engineering.

There are a number of resources for teaching a quantum computer programming course. Among them are \cite{Qiskit,hundt2022quantum,johnston2019programming,ying2024foundations}. These materials can be introduced to the students. 
Our emphasis during this course is practical teaching of three frameworks: Qiskit from IBM \cite{Qiskit}, PennyLane from Xanadu \cite{bergholm2018pennylane}, and Ocean from D-Wave \cite{Ocean}.

This paper proposes a six-step teaching approach for teaching quantum computer programming and quantum software engineering, which considers new advances in this field:
\begin{enumerate}
    \item Establishing the Quantum Foundation
    \item Quantum Software Engineering and Quantum Software Development Lifecycle
    \item Containerization for Quantum Computer Programming Education
    \item Programming with Qiskit Package
    \item Programming With PennyLane
    \item Programming With Ocean SDK
\end{enumerate}

Our teaching approach is based on the experience of teaching quantum computer programming and quantum software engineering in two countries. This approach suggests a common pattern that could eventually be generalized.



\section{Establishing the Quantum Foundation}
There are a number of resources for teaching quantum computing, such as \cite{Qiskit,mcmahon2007quantum,mermin2007quantum,nakahara2008quantum,nielsen2002quantum,bergholm2018pennylane,rieffel2011quantum} that can be introduced to the students who are interested to study and practice more about quantum computing concepts during the course.

Key topics in quantum computing such as superposition, entanglement, teleportation, or quantum algorithms, can be further discussed during some sessions by the lecturer.

We propose that the mentors explore the following foundational principles:

\begin{enumerate}
    \item Introduction to the fundamentals of quantum mechanics. 

    \item Description of qubit and quantum gates. 

    \item Teach the representation of quantum circuits. 

    \item Teach quantum algorithms such as quantum teleportation algorithm \cite{bennett1993teleporting}, Shor algorithm \cite{shor1994algorithms} for integer factorization, and the Grover search algorithm \cite{grover1996fast} and more.

    \item Exploration of quantum error correction and fault-tolerant quantum computing.

    \item Introduction to adiabatic quantum computing and quantum annealing.    
\end{enumerate}


\IEEEpubidadjcol

\section{Quantum Software Engineering and Quantum Software Development Lifecycle}

To equip students with the necessary skills for quantum computer programming, instructors should introduce a variety of quantum programming languages and frameworks. We suggest to introduce these three frameworks:

\begin{itemize}
    \item \textbf{Qiskit:} An open-source quantum computing framework developed by IBM \cite{Qiskit}.
    \item \textbf{PennyLane:} A platform that integrates quantum computing with machine learning, providing tools for hybrid quantum-classical computations \cite{bergholm2018pennylane}.
    \item \textbf{Ocean:} is a suite of open-source Python tools provided by D-Wave and accessible via the \textbf{Ocean} SDK \cite{Ocean}. It is useful for developing and running quantum annealing algorithms on D-Wave quantum annealers \cite{johnson2011quantum}. 
\end{itemize}

 Of course there are other options that can be explored by the mentors like:

\begin{itemize}
    \item \textbf{Microsoft Quantum Development Kit (QDK):} A comprehensive set of tools for quantum programming, including the Q\# programming language. Q\# is The quantum programming language provided by Microsoft's Quantum Development Kit (QDK), designed specifically for expressing quantum algorithms\cite{QsSpec2020}.
    \item \textbf{Intel Quantum Software Development Kit (SDK):} Intel's suite of tools designed for developing quantum applications and interfacing with their quantum hardware.
    \item \textbf{Cirq:} Cirq is An open-source framework designed by Google for creating, simulating, and running quantum circuits \cite{cirq_developers_2023_10247207}.
\end{itemize}

Various quantum software development frameworks, different lifecycle proposals, and an introduction to the concept of Quantum Software Engineering and its basic principles should be discussed during the lectures.

In addition to introducing quantum programming languages like Qiskit, PennyLane, DWave, Microsoft Quantum Development Kit, and Intel Quantum Software SDK, instructors can delve deeper into the quantum software engineering and quantum software development lifecycle we explored in \cite{dwivedi2024quantum}. The software development life cycle (SDLC) of hybrid classic-quantum applications consists of a comprehensive and multi-dimensional process \cite{stirbu2023full} that needs to be explored by the instructors. 


\section{Containerization for Quantum Computer Programming Education}

For instructors, effectively teaching quantum programming requires not only a strong grasp of quantum mechanics but also practical strategies to bridge the gap between classical and quantum domains \cite{romero2023enabling, romero2022using}. We propose that one such strategy involves the use of containerization technology for quantum computer programming, particularly Docker, as a preparatory step before introducing practical quantum programming.
Mentors can start with topics like understanding containerization, and then explore benefits of Docker for quantum programming. For this purpose they can explain Consistency and Portability, Isolation, and Ease of Setup of Docker Containers.
Here our proposed teaching strategy for this section is as follows:

\textbf{1. Introduction to Docker:}
Before diving into quantum programming, instructors should dedicate a session of the course to familiarizing students with Docker. This introduction should cover:
\begin{itemize}
    \item Basic concepts of containerization and Docker.
    \item Installation and setup of Docker.
    \item Creating, running, and managing Docker containers.
    \item Using Docker Hub to access and pull pre-built images.
\end{itemize}

\textbf{2. Practical Exercises with Docker and Transition to Quantum Programming:}
To solidify students' understanding, practical exercises should be integrated, such as:
\begin{itemize}
    \item Building a Docker container for a classical application.
    \item Managing containerized applications using commands.
    \item Exploring Dockerfile syntax to create custom containers.
    \item Setting Up Quantum Environments with Docker: Utilizing Docker to create environments for different quantum frameworks like qiskit, PennyLane, and D-Wave. Instructors can provide Docker images pre-configured with these frameworks, ensuring a smooth transition from theory to practice.
\end{itemize}

 By incorporating Docker into the quantum computer programming curriculum before introducing practical quantum programming concepts, instructors can provide a more efficient learning experience. This approach ensures that students have the necessary skills to manage complex software environments, and also have access to consistent and isolated environment during the course, allowing them to focus on mastering quantum computer programming concepts and techniques. Based on the level of class, mentors can also teach orchestration using Kubernetes for quantum that we explored in \cite{vlad2024} (called Qubernetes) in this stage of teaching. Another topic that is necessary to teach in this step is the concept of Quantum Serverless. We propose that the instructors explain quantum serverless in this stage.


\section{Programming with Qiskit Package}

After establishing a quantum foundation, we explain Qiskit programming, and how to prepare programming environments for qiskit. 
First, we need to discuss how to prepare the environment for programming using Qiskit SDK and simulations using Qiskit backends.
We also explain how to submit the jobs to the real quantum devices, e.g., using IBM real quantum computers. To give students a broader perspective, it's beneficial to introduce them to other quantum devices, such as Finland's HELMI quantum computer. By doing so, students learn to adapt their skills to different platforms and understand the diversity in quantum hardware. This can be achieved through:

\begin{enumerate}
    \item \textbf{Introduction to QPU (e.g., HELMI):} Instructors can provide their own local quantum computer examples. For HELMI, we provide an overview of Finland's HELMI quantum computer, its specifications, and its unique features compared to IBM's quantum computers. 
    \item \textbf{Access and Account Setup:} Explain the process of obtaining access to different QPUs, including any required account creation and credentials. Usually QPUs are available free for teaching purposes for the universities.
    \item \textbf{Quantum Programming Frameworks:} Discuss the specific quantum programming frameworks or APIs used to interact with QPUs. For example HELMI can work with Qiskit codes.
    \item \textbf{Job Submission Process:} Demonstrate the steps to submit a job to the QPUs:
    \begin{itemize}
        \item Writing a quantum program compatible with HELMI's framework.
        \item Selecting the HELMI quantum device for execution.
        \item Submitting the job and monitoring its progress.
        \item Retrieving and interpreting the results.
    \end{itemize}
    \item \textbf{Practical Examples:} Provide examples and exercises where students can write and submit quantum jobs to real quantum computers, reinforcing their learning and showcasing the variability between different quantum platforms.
\end{enumerate}


\section{Programming With PennyLane}

In this step of the proposed training, we explore PennyLane in depth. PennyLane is an open-source software framework that can be used basically for quantum computing, quantum machine learning and also quantum chemistry. PennyLane can be run on different hardware.

\subsection{Preparing Programming Environment for PennyLane}

To effectively teach students how to program using the PennyLane package, we recommend the following steps:

\begin{enumerate}
    \item \textbf{Prepare the Programming Environment}:
\textbf{Docker Container Setup}: We suggest that mentors guide students in setting up a Docker container specifically for PennyLane. This setup will ensure a consistent and controlled environment, which is crucial for avoiding compatibility issues and streamlining the learning process, as discussed in previous sections.

    \item \textbf{Utilize the Xanadu Codebook}:
    \begin{itemize}
        \item \textbf{Follow the Codebook}: The Xanadu website offers a comprehensive codebook\footnote{https://codebook.xanadu.ai/} for teaching PennyLane \cite{albornoz2023xanadu}. We advise lecturers to adhere to this codebook as it provides a structured and detailed curriculum for teaching the package effectively.
        \item \textbf{Incorporate Codercises}: The codebook includes numerous exercises, referred to as "codercises," designed to reinforce the concepts learned. Solving these exercises in class will provide students with hands-on practice, thereby enhancing their understanding and proficiency with PennyLane. Some of the topics on how to program using PennyLane that we suggest mentors teach them from this codebook are: Qubits, quantum circuits and gates; phase, rotations, and universal gate sets; Quantum state preparation; projective measurements, and observables; multi-qubit systems, entanglement, and Bell states; basic quantum algorithms like quantum teleportation, Hadamard transform, and the Deutsch Jozsa algorithm, Grover’s algorithm, The Quantum Fourier Transform (QFT), Quantum Phase estimation (QPE), Shor’s Algorithm. Obviously, there is more content like quantum machine learning, Hamiltonian simulation, VQE, QAOA, and Quantum Kernel that the mentors can decide to teach based on the course topics and class level.

    \end{itemize}
\end{enumerate}
By following these recommendations, students will have a well-prepared environment and a structured learning path, leading to a more effective and engaging educational experience with PennyLane.

\section{Programming With Ocean SDK}
We propose the following sequence for teaching Ocean SDK to be able to execute programs in D-Wave QPUs.
\subsection{Preparing the Programming Environment }
Start by setting up the programming environment for Ocean SDK. The Ocean Tools software is essential for running any program on the D-Wave Quantum Processing Unit (QPU). 

\subsection{Creating an Account }
Next, guide the students through the process of creating an account on the D-Wave website. An account is necessary to access the tools and resources provided by D-Wave and send the jobs to the DWave quantum computers.

\subsection{Using D-Wave Ocean Tools   }
Once the account is set up, explain how to use D-Wave Ocean Tools. Ocean’s SDK allows for Leap authentication, which is required to interact with the D-Wave QPU. A Solver API (SAPI) token from the Leap account is necessary to submit problems to Leap solvers.  

\subsection{Submitting Jobs to Quantum Devices    }
Teach how to submit jobs to D-Wave's real quantum devices. This involves using the SAPI token to authenticate and interact with the quantum solvers.   

\subsection{Connecting to Ocean SDK in Google Colab    }
Additionally, show how to connect to the D-Wave Ocean SDK within Google Colab. This enables a cloud-based approach to programming.   

\subsection{Understanding D-Wave Workflow     }
Discuss the workflow steps in D-Wave quantum computing. This includes:

\begin{itemize}
    \item \textbf{Formulating Problems}: Explain how to formulate problems as objective functions (cost functions).
    \item \textbf{Finding Solutions}: Teach how to find solutions through sampling.
\end{itemize}
   
\subsection{Solving Examples in different research areas using DWave quantum computers    }
Work through numerous examples across different research topics. This practical approach involves sending jobs to the QPU and analyzing the responses. This hands-on practice helps students become comfortable with programming on D-Wave quantum computers.

By following this sequence, students will gain a comprehensive understanding of the Ocean SDK and D-Wave quantum computing, enabling them to tackle complex problems using quantum algorithms.


\section{Conclusion and future work}

Our curriculum, which was developed collaboratively between university lecturers in Finland and Spain, integrates a comprehensive range of tools and methodologies, including containerization with Docker, Qiskit, PennyLane, and Ocean SDK. We discussed a new angle on teaching practices and methods for quantum software engineering and quantum computer programming. Since technology evolves rapidly, and changes in platforms should be considered to ensure that the most updated content is presented to graduates, this proposal is subject to review at regular intervals.

\subsection{Structuring the Curriculum}
The amount of material being taught is carefully aligned with the number of teaching hours and the duration of the course. This ensures that students can thoroughly understand and engage with the material without feeling overwhelmed. We emphasize a balanced approach to teaching, combining theoretical foundations with practical, hands-on experience.

\subsection{Unique Approach}
To the best of our knowledge, our curriculum structure is distinct from most existing courses. We have yet to encounter a course on quantum computer programming that offers and discusses three different SDKs: D-Wave, Qiskit, and PennyLane. This unique structure offers several strengths and advantages:

\begin{enumerate}
    \item Comprehensive Coverage: By presenting multiple frameworks, we ensure a well-rounded education. This approach allows students to understand the strengths and limitations of each SDK.
    \item Fair Presentation: As lecturers, it is essential to present a fair and balanced view of the available tools. While Qiskit is a prominent framework, PennyLane and D-Wave have unique values and applications that are equally important to explore.  
    \item In-depth Discussion: PennyLane’s integration with machine learning frameworks and D-Wave’s focus on optimization problems provide students with diverse perspectives on quantum computer programming applications.  
    \item Strengths of Qiskit, PennyLane, and Ocean: Despite presenting multiple frameworks, we acknowledge the robustness and versatility of Qiskit, Ocean, and PennyLane, which make them among the best tools for quantum computer programming education.
\end{enumerate}

\subsection{Future Works}
In the future, we plan to compare our approach to other methodologies for teaching quantum computer programming and quantum software engineering. This comparative analysis will help us determine if our approach is superior to existing methods. Additionally, we intend to develop and administer a detailed questionnaire to gather feedback from students regarding the effectiveness of our teaching strategies and the comprehensiveness of the curriculum. This feedback will be crucial in refining and improving our courses.

Our paper would greatly benefit from explicitly making the teaching materials available to a wider audience. Sharing our curriculum, lecture notes, and practical exercises can provide valuable resources for other educators and institutions.

By continuously evaluating and improving our teaching methods, we aim to provide the highest quality education in quantum computer programming and quantum software engineering. Our goal is to inspire and equip the next generation of quantum computing professionals with the knowledge and skills necessary to drive innovation in this exciting field of quantum computer programming and quantum software engineering.

Next on our agenda, we intend to distribute a questionnaire and, using the students' evaluations, present our findings. Additionally, we plan to conduct an empirical evaluation and publish the preliminary results.

\section*{Acknowledgement}
The authors appreciate Xanadu company and PennyLane Team.


\bibliographystyle{IEEEtran}
\bibliography{refs}

\end{document}